\documentclass[10pt,journal,compsoc]{IEEEtran}

\usepackage[T1]{fontenc}
\usepackage[utf8]{inputenc}

\usepackage{booktabs} 

\usepackage{layouts} 

\usepackage{xspace}
\newcommand\eg{e.g.\@\xspace} 
\newcommand\ie{i.e.\@\xspace}
\newcommand\wrt{w.r.t.\@\xspace}

\newcommand\cf{cf.\@\xspace}

\newcommand{\mytilde}{\raise.17ex\hbox{$\scriptstyle\mathtt{\sim}$}}

\usepackage[shortcuts]{extdash}
\usepackage{url}
\usepackage{subfig}
\usepackage{wrapfig}
\usepackage{units}
\usepackage{algorithm}
\usepackage{algpseudocode}
\usepackage{listings}
\usepackage{varwidth} 

\algnewcommand\algorithmicforeach{\textbf{for each}}
\algdef{S}[FOR]{ForEach}[1]{\algorithmicforeach\ #1\ \algorithmicdo}
\algrenewcommand{\algorithmiccomment}[1]{\hfill\texttt{//#1}}
\algrenewcommand{\algorithmicreturn}{\State \textbf{return}}
\algnewcommand{\False}{\textbf{false}}
\algnewcommand{\True}{\textbf{true}}

\algblockdefx{WHEN}{ENDWHEN}[1]%
  {\textbf{when }#1 \textbf{do}}%
  {\textbf{end do}}

\usepackage[shortlabels]{enumitem}
\setitemize{    leftmargin=0.4cm, noitemsep, nolistsep, topsep=0pt, parsep=0pt, partopsep=0pt}
\setenumerate{  leftmargin=0.4cm}
\setdescription{leftmargin=0.4cm}
\setlist{       itemsep=0pt, topsep=0pt, parsep=0pt}

\usepackage{color}

\usepackage{amsmath}
\usepackage{amsfonts}
\usepackage{amssymb}
\usepackage{mathtools}
\makeatletter
\newcases{lrdcases}
  {\quad}
  {$\m@th\displaystyle{##}$\hfil}
  {$\m@th\displaystyle{##}$\hfil}
  {\lbrace}
  {\rbrace}
\newcases{lrdcases*}
  {\quad}
  {$\m@th\displaystyle{##}$\hfil}
  {{##}\hfil}
  {\lbrace}
  {\rbrace}

\usepackage{cleveref} 
\crefname{section}{Section}{Sections}
\crefname{figure}{Figure}{Figures}
\crefname{table}{Table}{Tables}
\crefname{equation}{Eq.}{Eqs.}
\crefname{algorithm}{Algorithm}{Algorithms}

\makeatletter
\newcommand*\wt[1]{\mathpalette\wthelper{#1}}
\newcommand*\wthelper[2]{%
        \hbox{\dimen@\accentfontxheight#1%
                \accentfontxheight#11.3\dimen@
                $\m@th#1\widetilde{#2}$%
                \accentfontxheight#1\dimen@
        }%
}
\newcommand*\accentfontxheight[1]{%
        \fontdimen5\ifx#1\displaystyle
                \textfont
        \else\ifx#1\textstyle
                \textfont
        \else\ifx#1\scriptstyle
                \scriptfont
        \else
                \scriptscriptfont
        \fi\fi\fi3
}
\makeatother

\ifCLASSOPTIONcompsoc
  \usepackage[nocompress]{cite}
\else
  \usepackage{cite}
\fi

\usepackage[pdftex]{graphicx}

\begin{document}

\title{Agent-based Constraint Solving for Resource Allocation in Manycore Systems}

\author{
  Volker Wenzel$^\dagger$, Lars Bauer$^\dagger$, Wolfgang Schr\"oder-Preikschat$^\ddagger$, J\"org Henkel$^\dagger$\\
  $^\dagger$Chair for Embedded Systems (CES), Karlsruhe Institute of Technology (KIT), Germany\\
  $^\ddagger$ Friedrich-Alexander-Universit\"at (FAU), Erlangen-N\"urnberg, Germany\\
  lars.bauer@kit.edu, wosch@cs.fau.de, henkel@kit.edu
}

\IEEEtitleabstractindextext{
\begin{abstract}
  For efficiency reasons, manycore systems are increasingly heterogeneous, which makes the mapping of complex workloads a key problem with a high optimization potential.
  Constraints express the application requirements like which core type to choose, how many cores to choose, exclusively or non-exclusively, using a certain core, etc.
  In this work, we propose a decentralized solution for solving application resource constraints by means of an agent-based approach in order to obtain scalability.
  We translate the constraints into a Distributed Constraint Optimization Problem (DCOP) and propose a local search algorithm RESMGM to solve them.
  For the first time, we demonstrate the viability and efficiency of the DCOP approach for heterogeneous manycore systems.
  Our RESMGM algorithm supports a far wider range of constraints than state-of-the-art, leading to superior results, but still has comparable overheads \wrt computation and communication.
\end{abstract}
}

\maketitle

\IEEEraisesectionheading{
\section{Introduction and Related Work}
\label{sec:introduction}
}
\IEEEPARstart{T}{he} amount of cores per CPU is growing and expected to be around hundreds or
even thousands in the next few years \cite{borkar07thousands, IntelKnightsLanding2016}.
This creates the challenge of how to efficiently use the computational resources
of such large manycore systems, \ie, which computational resources should be assigned to which application.
The search space of the underlying problem grows excessively with the amount of available resources.
Highly dynamic workloads make it difficult to pre\-/determine the resource
allocation and call for \emph{runtime} resource management approaches.

\vfill
Among others, the authors of \cite{ebi_tape, cao2002arms, Ge:2010:DTM:1837274.1837417}
argue that centralized control of a huge amount of on\-/chip resources is to
become a bottleneck and that decentralized or \emph{agent\-/based}
resource management schemes have to be devised.
By \emph{agents}, we understand per\-/application software entities that negotiate the
resource requirements thereof.
Agent\-/based approaches to system management like~\cite{ebi_tape,kobbe_distrm}
are favorable for a multitude of reasons with scalability to bigger system sizes
being the main advantage.
Large manycore systems could theoretically be programmed by giving the
application programmer full flexibility in choosing computational resources (\eg cores) for the individual threads, but that would burden the application programmer with a great deal of additional intricacy.
These applications would be tailored to a specific manycore system, and
it is unclear how several applications would share the resources on the same
manycore system with acceptable resource allocation tradeoffs.
Alternatively, application programmers could remain agnostic of the
concrete computational resource on which a thread is executed, \ie,
resource management would lie entirely in the responsibility of the resource
management system.
This approach is also suboptimal in complex scenarios, \eg, with heterogeneous
computing resources, since the resource management system has no means to
differentiate more and less suitable resource allocations.

\begin{figure}[t]
  \begin{center}
    \includegraphics[width=\columnwidth]{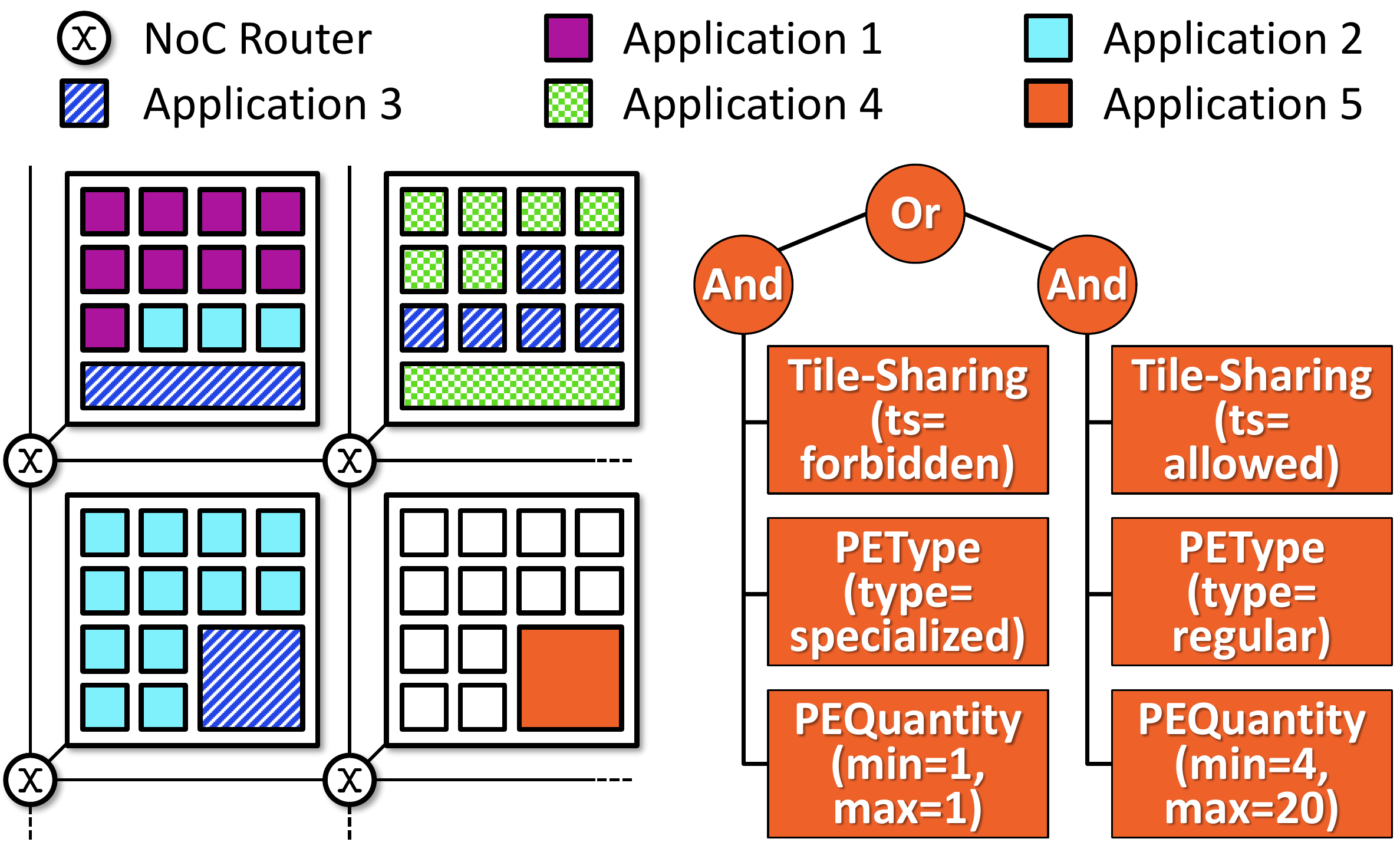}
  \end{center}
  \vspace{-1em}
  \caption{
    Example architecture of a tiled manycore system (left) and the constraint
    for Application~5 (right), showing that it needs either one specialized core
    exclusively (\ie, no other application may use any resource in the same tile),
    or several regular cores.
    Every application has a constraint of this form and the resource management needs to resolve resource requests at runtime.
    }
  \label{fig:tilearch}
\end{figure}

\newpage
We strongly believe that applications will have to be \emph{resource\-/aware}
\cite{hannig_resource_aware} and provide \emph{constraints}, \ie, preferences
for a desired resource assignment, to the resource management system.
By introducing support for constraints on resource assignments,
we give the application programmer freedom to influence the resource
assignment of an application.
For instance, constraints encompass a minimum and maximum amount of
required compute cores, on\-/chip memory or NoC\-/bandwidth necessary to
execute the application.
A concrete example for such a constraint is given in \cref{fig:tilearch} (right), where Application~5 is mapped to the small manycore architecture (left).
The application provides two implementation alternatives for its functionality and describes them as a hierarchical composition using the \texttt{Or} and \texttt{And} constraints.
One implementation can be executed on an exclusively\-/reserved tile (\texttt{Tile-Sharing} forbidden) using a single (\texttt{PEQuantity}) specialized (\texttt{PEType}) core.
Alternatively, a larger amount of regular cores scattered over several tiles of the manycore system may be used.
Existing examples of applications that benefit from using constraints to express
changing resource demands during different phases of their execution comprise
applications that use multigrid methods \cite{bungartz_multigrid}.

Surveys of state\-/of\-/the\-/art manycore resource allocation systems are given in \cite{singh_review, Singh:2017:SCS:3071073.3057267}.
Our contribution in the presented taxonomy in~\cite{singh_review} lies in the
branch of \emph{distributed} resource management systems that operate at
\emph{runtime} and support \emph{heterogeneous resources}.
A similar NoC\-/based system architecture is targeted in \cite{Fattah:2013:SHC:2463209.2488782}, but it only employs a centralized stochastic hill climbing approach for resource management in order to find a contiguous mapping of resources.
The approaches in \cite{kobbe_distrm, 6654651} are distributed, but they only support a small subset of the constraints conceivable in our approach.
The \emph{Distributed Resource Manager} (DistRM) \cite{kobbe_distrm} foresees one agent per application and features local decision\-/making.
Moreover, DistRM also supports \emph{malleable} applications, \ie, applications that can dynamically adapt their resource usage at runtime.
The more recent decentralized resource management scheme introduced in \cite{anagnostopoulos_distrmclone} elaborates on \cite{kobbe_distrm, 6654651}.
This work has the most similar objective to our work, and we therefore compare our proposed technique to it in \cref{sec:evaluations}.
It introduces a workload\-/aware distributed runtime framework for malleable applications, \ie, applications that support changing their resources at runtime.
In contrast to RESMGM, \cite{anagnostopoulos_distrmclone} does not support the same variety of constraints and there is no straight forward potential to extend it.

Agent\-/based approaches for resource management can be constructed in multiple ways.
A great body of literature borrows from market inspired principles.
Examples are \cite{ebi_tape} and \cite{melissaris_agora}, which is a distributed resource management based on market principles that proactively balances reliability degradation.
Typically these approaches have relatively low overheads, but an uncontrollable solution quality.
Another branch of agent\-/based resource managers employs methods from mathematical optimization.
An example is \cite{wildermann_multi_objective}, which is a runtime resource management approach that is based on nonlinear programming.
It allows a distributed application embedding for dealing with soft system\-/wide constraints.
Typically these approaches come with high overheads, but allow trading off solution quality with computational and message passing overheads.
An orthogonal aspect of resource allocation schemes is, whether they work at design time, runtime, or both (hybrid).
An example of a hybrid scheme is presented in \cite{olsen_res_mgmt}, which is a performance-aware resource management scheme for many-core architectures.
Note that \cite{olsen_res_mgmt} performs application mapping in a centralized way.

\vspace{0.5em}\noindent
The \textbf{novel contributions} of this paper are as follows:
\begin{itemize}
\item   We are the first to tackle the problem of \emph{manycore resource management}
        with an \emph{agent\-/based constraint solving} approach to the best of our
        knowledge.
\item   For the first time, we formally translate the problem of agent\-/based
        solving of resource constraints for resource\-/aware applications in
        manycore systems into a versatile mathematical problem formulation: the
        \emph{Distributed Constraint Optimization Problem} (DCOP).
        This yields a new approach of a fully decentralized constraint solving
        for resource\-/aware applications.
\item   We propose RESMGM, an optimized DCOP algorithm for
        decentralized resource management.
        It uses the maximum gain message (MGM) communication \cite{mgm} along with a multitude of
        novel heuristics that improve execution time and solution quality significantly.
        Among those heuristics are the \emph{Smart initialization heuristic}
        to incorporate the respective system state, the \emph{Early
        termination heuristic}, which adds improved termination detection,
        and the \emph{Incorporating system information
        heuristic}, which exploits the architecture of the manycore system in
        order to make the evaluation of constraints more efficient.
\end{itemize}

\section{Problem Formulation and System Overview}
\label{sec:problem_formulation}

\subsection{System Overview}
\label{sec:sys_overview}

\begin{figure}[t]
  \begin{center}
    \includegraphics[width=0.95\columnwidth]{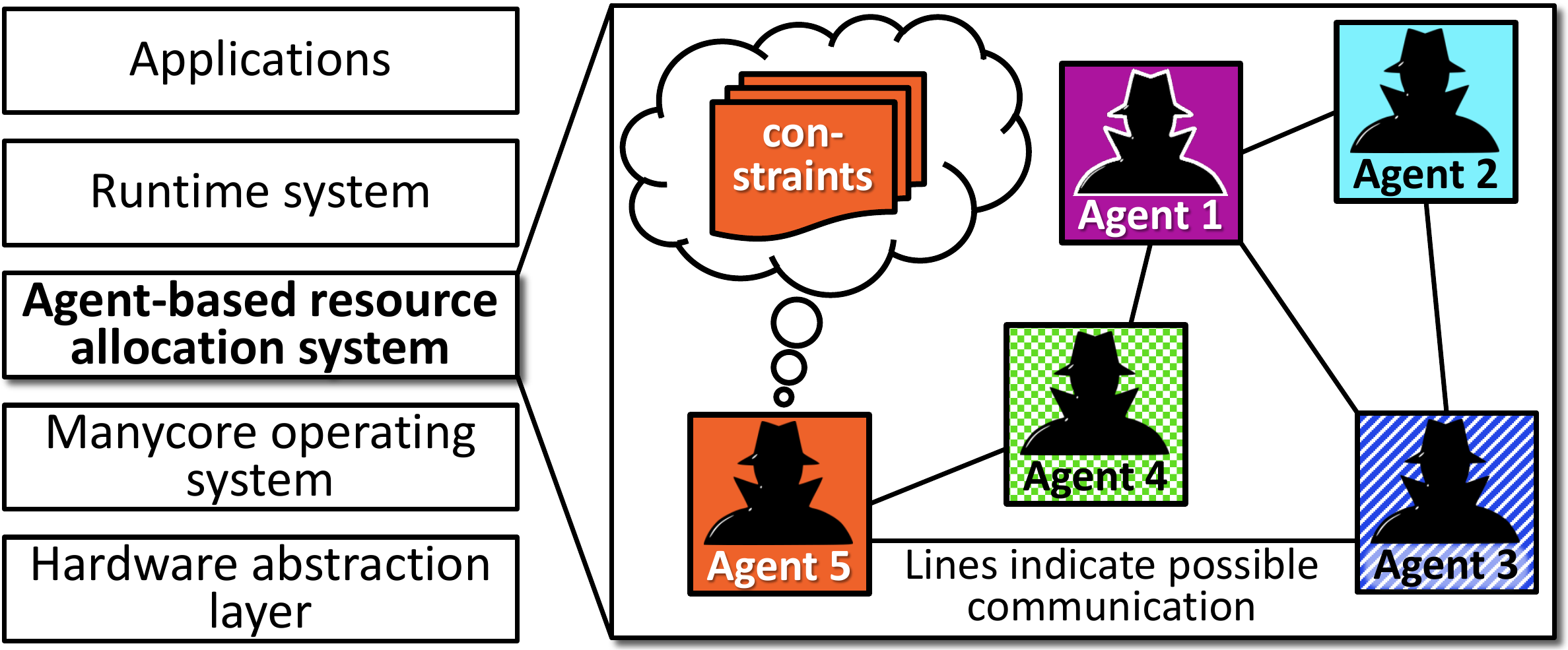}
  \end{center}
  \vspace{-1em}
  \caption{
    Overview of targeted technology stack (left) and agent interactions (right).
    The agent\-/based resource management is implemented as a part of a manycore operating system.
    }
  \label{fig:agent_cartoon}
  \vspace{0.5em}
\end{figure}

It is expected that future manycore CPUs will consist of a big amount of so
called \emph{compute tiles} with a tiled \emph{Network\-/on\-/Chip} (NoC)\-/based
communication infrastructure.
By \emph{compute tile} we understand the basic unit of future manycore systems,
consisting of one or several compute cores, caches, and a network adapter to
interface the NoC.
A typical software stack for manycore systems is
illustrated on the left side of \cref{fig:agent_cartoon}.
The \emph{agent\-/based resource management system}, the focus of this publication, is situated on top of the operating system and below the application and its runtime system (providing standard library calls).
By an \emph{agent} we understand a software entity that exclusively negotiates the
resource demands of its application, \ie, there is one agent per application.
Agents can communicate with other agents via message passing.
An example of the agent's interactions is shown on the right side of \cref{fig:agent_cartoon}.
An application that demands new or different resources, passes a \emph{constraint} along with its request for resources to the agent system via the agent-system API.
The agents then start a distributed constraint solving optimization that will be described in \cref{subsec:dcop_formal}.
Resource requests that arrive before the constraint solving finishes will be buffered and processed in the subsequent optimization.
This ensures termination of the started resource management process, instead of aiming at `a moving target'.
Afterwards, all buffered requests (potentially from different applications) are processed together.

\subsection{Problem Formulation}
\label{subsec:problem_def}

In the following, a formal problem formulation is provided.
We assume the manycore system to consist of \emph{allocatable resources} $r_e$
\begin{equation}
  r_e \in R = \{r_1,\ldots,r_r\}
\end{equation}
where $R$ is the set of all resources, and $r$ is the total number of resources.
The resources could be computational resources (\ie cores), chunks of scratchpad
memory, NoC bandwidth or other countable resources present in the manycore
system.
For the sake of simplicity, we will limit our discussion to compute cores.
Every resource is part of a \emph{compute tile} $t_m$.
\begin{equation}
  t_m \in T = \{t_1,\ldots,t_t\}
\end{equation}
\begin{equation}
  tile : R \mapsto T
\label{eqn:tilemap}
\end{equation}
The function $tile$ maps resources to their corresponding compute tile, \eg, $tile(r_e)=t_m$.
$H$ is the set of heterogeneous resource types.
In our case there are three different computational resource types, namely
$
H = \big\{ \mathrm{REGULAR},\ \mathrm{RECONFIG},\ \mathrm{STREAM} \big\}
\label{eqn:het_types}
$, where $\mathrm{REGULAR}$ may represent a standard general purpose processor, 
$\mathrm{RECONFIG}$ may represent a runtime reconfigurable processor, and 
$\mathrm{STREAM}$ may represent a dedicated stream processing unit. 
Note that our approach is not limited to specific resource types or a specific
number of different resource types.
The function
$
  type : R \mapsto H
  \label{eqn:typemap}
$ maps resources to their corresponding heterogeneous resource type, \eg, $type(r_e)=\mathrm{RECONFIG}$.

The set of applications is given by
$
  A = \{a_1,\ldots,a_p\} \cup \{a_\textrm{new}\}
$
We refer to $\{a_1,\ldots,a_p\}$ as \emph{a priori} applications with $p$ being the number of a priori applications.
$a_\textrm{new}$ ($\equiv a_{p+1}$) represents a new application for which resources shall be found.
Note that our approach is not limited to resource allocations for a single new application.
It also supports multiple new applications and applications that change their resource requirements at runtime, but for the sake of simplicity, we limit our discussion to having just one new application.
The function $assigned$ provides the current resource assignment of an application
\begin{equation}
  assigned : A \mapsto \mathcal{P}(R)
\end{equation}
where $\mathcal{P}(R)$ denotes the power set of the available resources, \ie, the resource assignment of an application is a subset of all resources.
Initially $assigned(a_\textrm{new}) = \emptyset$.

All applications $a_j \in A$ provide a \emph{soft, valued} constraint,
\ie, a constraint that can be partially fulfilled.
\begin{equation}
C_j : \mathcal{P}(R) \mapsto \mathbb{R}^+_0 \cup \{\infty\}
\label{eqn:def_constraint}
\end{equation}
We assume one constraint $C_j$ per application $a_j$ that takes a set of resources as input and evaluates them \wrt the constraint.
Note that a constraint may internally consist of sub\-/constraints.
For instance the example shown in \cref{fig:tilearch} (right) shows actually \emph{the} (one) constraint for Application~5, describing its resource requirements.
The constraints and their internal composition are formalized further in \cref{sec:methods,eqn:constraint_and,eqn:constraint_or,eqn:constraint_union}.
The problem to be solved is to find a resource assignment
for $a_\textrm{new}$ that minimizes a
\emph{cost function}, given by
\vspace{+0.2em}
\begin{equation}
cost : \mathcal{P}(R)^{\,p+1} = \mathcal{P}(R) \times \ldots \times \mathcal{P}(R) \mapsto \mathbb{R}^+_0 \cup \{\infty\}
\vspace{+0.2em}
\end{equation}
It takes the $p+1$ resource assignments of the $p$ a priori applications and the new application as input, evaluates their constraints and calculates the corresponding cost for the resource assignments,
where $cost\!=\!\infty$ means that at least one hard constraint is violated.
The cost function returns smaller values for resource allocations
that are more useful (\ie, have a higher \emph{utility}) to the application.
The goal is to \textbf{minimize} $cost$ by finding a well\-/suited resource
assignment.
This is to be done by the agents, whose local knowledge is only sufficient to
evaluate some of the terms in $cost$.
The cost function is calculated as
\vspace{+0.2em}
\begin{equation}
  cost\!=\!\sum_{j=1}^{p+1} \Big(C_j\big(assigned'(a_j)\big)\Big) + \sum_{j=1}^{p+1} \Big(C_{\textrm{migr},j}\Big) + C_\mathrm{Sys}
  \label{eqn:cost_func_explicit}
\end{equation}
where the function $assigned'$ provides the new (to be examined) resource assignment of an application.
Depending on the configuration of the resource management system, it is
permissible that the a priori applications change their resource assignment from $assigned(a_j) \in \mathcal{P}(R)$ to $assigned'(a_j) \in \mathcal{P}(R)$
in order to allow the new application to obtain a better resource assignment
and/or to improve the overall system performance.
This incurs a migration cost represented by a set of migration constraints
\begin{equation}
C_{\textrm{migr},j} : \mathcal{P}(R)^2 \mapsto \mathbb{R}^+_0 \cup \infty
\end{equation}
where $C_{\textrm{migr},j} = \infty$ means that application $a_j$ does not support migration.
Lastly, it has to be made sure that resources are not assigned to different
applications at the same time.
We formally capture this with the system constraint $C_\mathrm{Sys}$, which we
introduce more in detail in \cref{eqn:systemconstraint}.

\section{Distributed constraint solving with the RESMGM algorithm}
\label{sec:methods}

\subsection{Formulating resource management as distributed constraint optimization problem}
\label{subsec:dcop_formal}
We first outline how the formulated problem (see \cref{sec:problem_formulation})
can be expressed as a \emph{Distributed Constraint Optimization Problem} (DCOP),
and explain how it can be applied to solve resource constraints for
manycore architectures.
Our definitions regarding DCOP are based on \cite{yeoh08bnbadopt}.
Every \emph{agent} represents one software \emph{application} in the system.
The set of all agents in the system is given by
\begin{equation}
  \tilde{A} = \{\tilde{a}_1,\ldots,\tilde{a}_p\} \cup \{\tilde{a}_\textrm{new}\}
\end{equation}
in analogy to the set of applications $A$.
Every agent in the system has exactly one \emph{resource assignment}, except for
$\tilde{a}_\textrm{new}$, for which it is yet to be determined.
The resource assignments of all agents make up the the \emph{resource allocation}.

Every agent $\tilde{a}_j$ maintains a \emph{variable} $v_j$ from the
\emph{domain} $D$ in which it records the agent's current local view of the
system's global resource allocation.
That is, $v_j$ contains the agent's local view about the resource owner for
all $r$ resources, \ie, itself, another agent, no agent (a free resource), or an
unknown status.
\begin{equation}
  \begin{split}
    v_j = ( \alpha_1,\ldots,\alpha_r ) \in D = d^{r} = d \times \ldots \times d\\
    \alpha_i \in d = \tilde{A} \cup \{\textrm{FREE}\} \cup \{\textrm{UNKNOWN}\}
  \end{split}
  \label{eqn:variable}
\end{equation}
Note that this local view is typically incomplete, \ie, it does not contain all
information of the global view and might even contain \emph{wrong} information temporarily.
Each agent $\tilde{a}_j$ has its own local view $v_j$ and
each agent must at least have knowledge of all its own
resources in $v_j$.
The set of all variables is given by
$
  V = \{ {v_1,\ldots,v_p}\} \cup \{v_{\textrm{new}}\}\\
$
The incomplete/wrong information will be incrementally updated by communicating
with other agents, as explained in the following.

Every agent strives to fulfill the constraint of its application as good as
possible, taking into account the other agents.
In analogy to \cref{eqn:def_constraint}, constraints are functions that rate the
quality of resource assignments.
The original definition of $C_j$ in \cref{eqn:def_constraint} expects the resources that are actually assigned to an application $a_j$ as input.
In the following, we use an extended version of $C_j$ that also accepts the local resource allocation view of an agent instead.
For instance, the constraint $C_j$ rates the local resource allocation view
$v_j$ of agent $\tilde{a}_j$ to inform how good the constraint is fulfilled
by that assignment.
The set of all constraints is given by:
\begin{equation}
  C = \{C_1,\ldots,C_p\} \cup \{C_{\textrm{new}}\}
\end{equation}
The agents examine different resource assignments (\ie intermediate assignments
to variable $v_j$), communicate that to other agents to identify and solve
conflicting assignments, and try to optimize the global cost function.
The cost function (lower is better) for intermediate variable assignments
$v_j$ is given by
\begin{equation}
  \begin{split}
    \wt{cost}: D^{\,p+1} = D \times \ldots \times D \mapsto \mathbb{R}^+_0 \cup \{\infty\}\\
    \wt{cost} = \sum_{j=1}^{p+1} \Big(C_j(v_j)\Big) + \sum_{j=1}^{p+1} \Big(C_{\textrm{migr},j}\Big) + C_\mathrm{Sys}
  \end{split}
  \label{eqn:cost_agent}
\end{equation}
in analogy to the definition in \cref{sec:problem_formulation}.
It is noteworthy to mention that global knowledge is not necessary in order to
minimize $\wt{cost}$.
The agents minimize $\wt{cost}$ by sending/receiving messages to/from other
agents, evaluating their constraints locally, and optimizing the assignment
of their own variables.

\begin{figure*}[t]
  \begin{center}
    \includegraphics[]{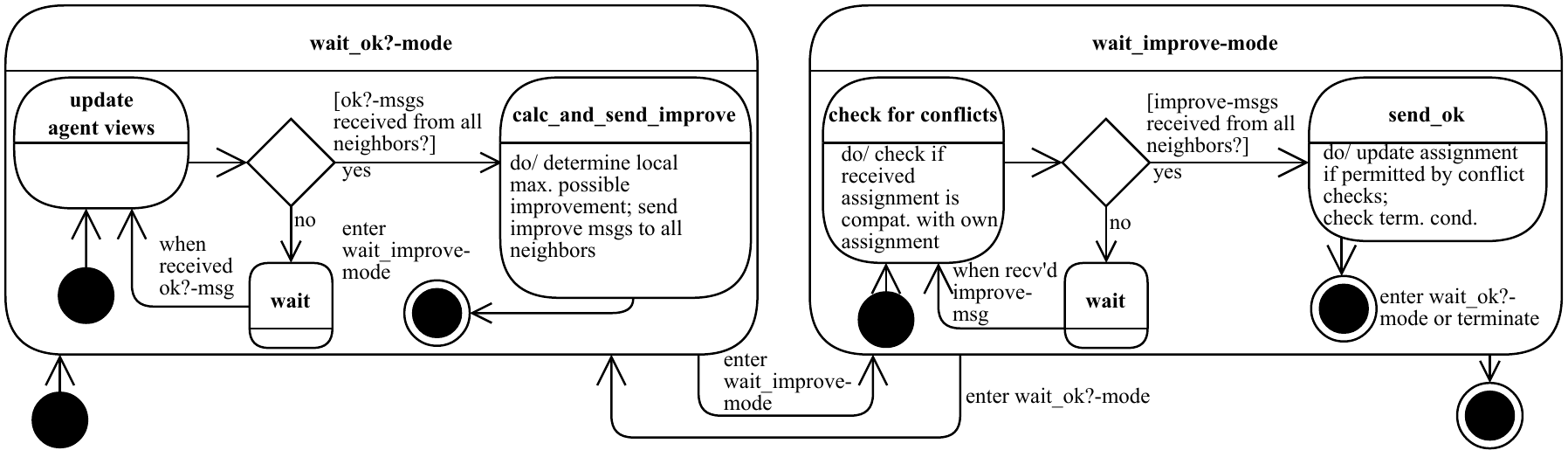}
  \end{center}
  \vspace{-1em}
  \caption{
    UML state machine of RESMGM.
    RESMGM alternates between \texttt{wait\_ok?}-mode and \texttt{wait\_improve}-mode until termination. Transition between states occurs, when all \emph{ok?}- or \emph{improve}-messages have been received.
    }
  \label{fig:sm_MGM}
\end{figure*}

Many different kinds of constraints are conceivable.
In the following, $\tilde{a}$ is the agent to which the constraint applies,
$v = (\alpha_1,\ldots,\alpha_r) \in D$ (cf.\ \cref{eqn:variable}) is a possible resource assignment for agent
$\tilde{a}$, and $v[r_e] = \alpha_e, r_e \in R$ is the local view of agent $\tilde{a}$ about the owner/status of resource $r_e$.
We envision the following constraints for our system:
$C_\mathrm{PEQuantity}$ constrains the number of cores in a resource assignment:
\begin{equation}
\begin{split}
  C_\mathrm{PEQuantity}(v, \tilde{a}, \textrm{minPEs}, \textrm{maxPEs}) = \\
  \begin{cases}
    0       & \textrm{minPEs} \leq \textrm{countPE}(v, \tilde{a}) \leq \textrm{maxPEs} \\
    \infty  & \textrm{else}
  \end{cases}
\end{split}
\end{equation}
\begin{equation}
  \textrm{countPE}(v, \tilde{a}) = \Big|\big\{v[r_e], r_e \in R : v[r_e]=\tilde{a}\big\}\Big|
\end{equation}
where minPEs, maxPEs $\in \mathbb{N}_0$ are the minimum and
maximum amount of resources in the resource assignment
($\textrm{minPEs} \leq \textrm{maxPEs}$), and countPE determines the amount of PEs assigned to $\tilde{a}$ in the examined resource assignment $v$.
$C_\mathrm{PEQuantity}$ evaluates to the constraint cost, \ie, whether or not the constraint is fulfilled.
Similarly, $C_\mathrm{PEType}$ constrains the resource types in a resource assignment:
\begin{equation}
C_\mathrm{PEType}(v, \tilde{a}, \textrm{PEtype})\!=\!
\begin{cases}
0       &
  \begin{aligned}
    & \forall\,r_e \in R, v[r_e] = \tilde{a} :\\
    & type(r_e) = \textrm{PEtype}
  \end{aligned} \\
\infty  & \textrm{else}
\end{cases}
\end{equation}
where $type(r_e) \in H$ is the resource type of $r_e$
(see \cref{eqn:het_types,eqn:typemap}) and $\textrm{PEtype} \in H$ is the type of the requested
resource.
The $C_\mathrm{TileSharing}$ constraint allows to request an exclusive access to tiles, \ie, no resource in any of the tiles of the requesting agent may be given to another agent.
Additionally, the resource management system should minimize the number of tiles
in the resource assignment.
\begin{equation}
C_\mathrm{TileSharing}(v, \tilde{a}) =
\begin{cases}
0      &
  \begin{aligned}
    & \forall\,r_e \in R, v[r_e] = \tilde{a}\ \ \nexists\,r_f \in R: \\
    & tile(r_e) = tile(r_f)\ \wedge\\
    & v[r_e] \notin \big\{ \tilde{a}, \mathrm{FREE} \big\}\\
  \end{aligned} \\
\infty & \textrm{else}
\end{cases}
\end{equation}
$C_\mathrm{Downey}$ provides a hint on the scalability of an application.
\begin{equation}
\begin{split}
  C_\mathrm{Downey}(v, \tilde{a}, \textrm{Dow}_\sigma, \textrm{Dow}_A) = \\
  \begin{cases}
    1 / f_{ \textrm{Downey} } & f_{ \textrm{Downey} }\big( \textrm{Dow}_\sigma, \textrm{Dow}_A, \textrm{countPE}(v, \tilde{a}) \big) \neq 0 \\
    \infty                    & else
  \end{cases}
\end{split}
\end{equation}
$f_{ \textrm{Downey} } \in \mathbb{R}^+_0$ is a formula to model the speedup
of parallel applications \cite{downey}.
The model depends on the average parallelism $\textrm{Dow}_A \in \mathbb{N}$,
and the parameter $\textrm{Dow}_\sigma \in \mathbb{R}^+$,
which are both provided by the application programmer.
In order to build more complex constraints, we introduce the conjunction and
disjunction of constraints and to allow for heterogeneous constraints, we define union.
\begin{equation}
  C_\mathrm{And}(v, C_1, C_2) = \textrm{max}\big\{C_1(v), C_2(v)\big\}
  \label{eqn:constraint_and}
\end{equation}
\begin{equation}
  C_\mathrm{Or}(v, C_1, C_2)  = \textrm{min}\big\{C_1(v), C_2(v)\big\}
  \label{eqn:constraint_or}
\end{equation}
\begin{equation}
  C_\mathrm{Union}(v, C_1, C_2)  = C_1(v) + C_2(v)
  \label{eqn:constraint_union}
\end{equation}
Here, $C_1$ and $C_2$ are independent constraints (\eg $C_\mathrm{PEQuantity}$, nested $C_\mathrm{And}$ etc., see \cref{fig:tilearch}).
For the And/Or constraint, $C_1$ and $C_2$ are combined by max/min to ensure that the worst/best resource assignment is considered.
The union constraint allows to request heterogeneous resources.
For instance, requesting regular and reconfigurable PEs is possible by creating two separate constraints (one for regular and one for reconfigurable with the corresponding PE quantities etc.) and then combining them using the union constraint.
In order to guarantee a consistent system state (\eg, there must not be two agents that want to use the same resource), we introduce the notion of a system constraint.
\begin{equation}
C_\mathrm{Sys}  =
\begin{cases}
0      &
  \begin{aligned}
    & \forall\,\tilde{a}_j, \tilde{a}_k \in \tilde{A}, \tilde{a}_j \ne \tilde{a}_k \ \ \nexists\,r_e \in R: \\
    & v_j[r_e] = \tilde{a}_j \wedge v_k[r_e] = \tilde{a}_k
  \end{aligned} \\
\infty & \textrm{else}
\end{cases}
\label{eqn:systemconstraint}
\end{equation}
The list of constraints presented here is by no means exhaustive and can be extended to capture other types of constraints according to the specific requirements of the applications.
However, for the sake of simplicity, we limit our considerations to the
constraints presented above.
The entire DCOP is given as a tuple of
$ 
  \textrm{DCOP} = \left\langle\,V,D,C,\tilde{A},\wt{cost},tile,R\,\right\rangle
  \label{eqn:dcop}
$.

\subsection{The RESMGM algorithm}
\label{subsec:alg_details}

A variety of algorithms exist to solve the DCOP formulation established in \cref{eqn:dcop}.
Preliminary tests show the \emph{Maximum Gain Message} (MGM) \cite{mgm} algorithm to be a well\-/suited candidate for the purpose of manycore resource management.
MGM is an agent\-/based local search algorithm.
It is \emph{incomplete}, \ie, it is not guaranteed to find an optimal solution, whereas complete DCOP algorithms, like BnB\-/ADOPT \cite{yeoh08bnbadopt}, showed poor scalability in our preliminary tests due to excessive overheads in computational effort and/or message passing.
Additionally, MGM comes with an \emph{anytime property}, \ie, it always improves on a partial solution that can be extracted if the algorithm needs to be interrupted before termination.

Starting from MGM as a baseline, we analyze its performance and behavior when applied to the specific scenario of manycore resource management (\cf \cref{subsec:dcop_formal}) and we conceive multiple novel heuristics that improve execution time and solution quality significantly, altogether building the foundation of our proposed \emph{Resource Management MGM} (RESMGM) algorithm.
An overview of the functioning of RESMGM is illustrated in \cref{fig:sm_MGM}.
A more in-depth understanding follows from the pseudo codes in \cref{alg:mgm_p1,alg:mgm_p2}\footnote{The pseudo-code has been split in two parts to improve readability.} and borrows its terminology from \cite{dba}.

The same algorithm is executed on every agent and communication takes place via asynchronous messages between \emph{neighboring} agents, \ie, agents that have a constraint between them (for instance the system constraint in \cref{eqn:systemconstraint}).
In RESMGM, message delivery is assumed to be reliable, but delays are tolerated.

Neighboring agents exchange two types of messages between each other:
Firstly the \emph{ok?}\-/message contains an agent's designation and the current assignment of its variables (see \cref{eqn:variable}).
The receiving agent adds its neighbors' current variable assignment to its $\mathit{agent\_view}$, which is a possibly incomplete and outdated record of its neighbors' resource assignment.
After having received the \emph{ok?}\-/messages from all neighbors, an agent uses its $\mathit{agent\_view}$ to calculate its current constraint cost and determines for which assignment of its variables it can minimize its constraint cost and by how much (=improvement).
Subsequently, the agent communicates its improvement via \emph{improve}\-/messages to all of its neighbors.
To avoid oscillations, the agent is only allowed to change its variable assignment, if its improvement is maximum among its neighbors.
Non\-/neighboring agents can change their variable assignments concurrently.

RESMGM alternates between two modes, \ie, the \texttt{wait\_ok?}\-/mode and the \texttt{wait\_improve?}\-/mode as illustrated in \cref{fig:sm_MGM}.
In the \texttt{wait\_ok?}\-/mode, the agent can either be waiting for all \emph{ok?}\-/messages to arrive from its neighbors or it can be waiting for all \emph{improve}\-/messages to arrive in the \texttt{wait\_improve?}\-/mode.
At the end of the processing in the \texttt{wait\_ok?}\-/mode, the agent sends out the \emph{improve}\-/messages and switches to the \texttt{wait\_improve?}\-/mode.
Similarly, at the end of the processing in the \texttt{wait\_improve?}\-/mode, the agent sends out \emph{ok?}\-/messages and switches to the \texttt{wait\_ok?}\-/mode.
RESMGM alternates between these two modes until termination.
The other agents execute the same steps.
Synchronization happens when waiting for all \emph{ok?}-/\emph{improve}-messages to arrive from the neighbors.

The algorithm improves the solution quality mono\-ton\-ously, \ie, if an agent's variable assignment does not change for any agent in a round, it will not change for any agent in the remaining rounds.
The \emph{ok?}\-/message has a Boolean field that encodes, whether the sending agent has changed its variable assignments in the last round or whether it has observed a change from any other agent's \emph{ok?}\-/message.
Due to the high connectedness of the constraint tree, it is safe to terminate RESMGM when one agent does not see any change from other agents in a round.

\begin{algorithm}
  \caption{ RESMGM -- \texttt{wait\_ok?}-mode}
  \label{alg:mgm_p1}
  \begin{algorithmic}[1]
    \State \textsc{initialize()} \label{alg:mgm_p1:init}
    \State \textsc{send\_ok()} \textrm{ //see \cref{alg:mgm_p2}. Note that it is easier to comprehend the algorithms by skipping the \textsc{send\_ok()} function for now and continuing reading the next lines, where the \textbf{ok?} messages are received from all neighboring agents.}
    \State \texttt{wait\_ok?}-mode: \label{alg:mgm_p1_wait_ok_mode}
    \WHEN{ \textbf{received} (\textbf{ok?},\enspace $\tilde{a}_n$,\enspace $v_n$,\enspace $change_n$)} \label{alg:mgm_p1:received:ok}
      \State $\mathit{counter} \leftarrow \mathit{counter} + 1$
      \State $\mathit{agent\_views} \leftarrow \mathit{agent\_views} \cup v_n$
      \If{ $\mathit{counter} = \mathit{number\_of\_neighbors}$ }  \label{alg:mgm_p1:allreceived}
        \State \textsc{calc\_and\_send\_improve()}
        \State $\mathit{counter} \leftarrow 0$
        \State \textrm{enter} \texttt{wait\_improve?}-mode \textrm{ //see \cref{alg:mgm_p2}}
      \EndIf
    \ENDWHEN
    \item[]

    \Procedure{ \textbf{calc\_and\_send\_improve()} }{}
      \State $\mathit{current\_cost} \leftarrow \wt{cost}(v_l \cup \mathit{agent\_views})$ //see \cref{eqn:cost_agent} for the definition of $\wt{cost}$;
      $\mathit{agent\_views}$ directly contains the local views of the neighboring agents, \ie, $\mathit{agent\_views} \in D \times ... \times D$\label{alg:mgm_p1:calc_cost_current}
      \State \begin{varwidth}[t]{\linewidth}
                $v'_l \leftarrow$ determine new variable assignment that gives \par
                the maximum improvement\label{alg:mgm_p1:new_variable_assignment}
             \end{varwidth}
      \State \begin{varwidth}[t]{\linewidth}
                $\mathit{my\_improve} \leftarrow current\_cost - \wt{cost}(v'_l \cup \mathit{agent\_views})$\label{alg:mgm_p1:calc_cost_new}
            \end{varwidth}
      \If { $\mathit{my\_improve} > 0$ }
        \State $\mathit{can\_move} \leftarrow \True$
      \Else
        \State $\mathit{can\_move} \leftarrow \False$
      \EndIf
      \ForEach {neighboring agent $\tilde{a}_n$}
        \State \begin{varwidth}[t]{\linewidth}
        \textbf{send} (\textbf{improve},\enspace $\tilde{a}_l$,\enspace $\mathit{my\_improve}$,\enspace $\mathit{current\_cost}$,
        $\mathit{my\_termination\_counter}$) to $\tilde{a}_n$ \label{alg:mgm_p1:send_improve}
        \end{varwidth}
      \EndFor
    \EndProcedure
    \algstore{break1}
  \end{algorithmic}
\end{algorithm}

\begin{algorithm}
  \caption{ RESMGM -- \texttt{wait\_improve?}-mode}
  \label{alg:mgm_p2}
  \begin{algorithmic}[1]
    \algrestore{break1}
    \State \texttt{wait\_improve?}-mode: \label{alg:mgm_p2:wait_improve_mode}
    \WHEN{ \begin{varwidth}[t]{\linewidth}
    \textbf{received} (\textbf{improve}, $\tilde{a}_n$, $v_n$, $\mathit{improve_n}$, $\mathit{eval}$,
    $\mathit{termination\_counter}$)}
    \end{varwidth}
      \State $\mathit{counter} \leftarrow \mathit{counter} + 1$
        \State \begin{varwidth}[t]{\linewidth}
        //Check whether the received variable assignment $v_n$ conflicts with our $v'_l$ (\eg if $\tilde{a}_n$ wants to use the same resource as we want).
        If that is the case and the improvement of $\tilde{a}_n$ is better than ours, then we stick with our old variable assignment.
        In case of a tie (identical improvements) a unique agent ID decides.
        \label{alg:mgm_p2:biggest_improvement_comment}
        \end{varwidth}
        \If { $\big(assignment\_conflict(v_n, v'_l)\big)\; \wedge
              \big(\,(\mathit{improve} > \mathit{my\_improve} \; \vee $
              $(\mathit{improve}=\mathit{my\_improve}$ $\wedge$ $UID(\tilde{a}_n)$ < $UID(\tilde{a}_l$))\,\big)}\label{alg:mgm_p2:biggest_improvement}
          \State $\mathit{can\_move} \leftarrow False$
        \EndIf
      \If { $\mathit{counter} = \mathit{number\_of\_neighbors}$ }
        \State \textsc{send\_ok()}
        \State $\mathit{counter} \leftarrow 0$
        \State clear $\mathit{agent\_views}$
        \State enter \texttt{wait\_ok?}-mode \textrm{ //see \cref{alg:mgm_p1}}
      \EndIf
    \ENDWHEN
    \item[]
    \Procedure{send\_ok()}{}
      \State \mbox{$\mathit{my\_termination\_counter} \leftarrow \mathit{my\_term'\_counter} + 1$}
      \If{$can\_move$}
      \State $v_l \leftarrow v'_l$\label{alg:mgm_p2:make_move}
      \State $change_l \leftarrow my\_termination\_counter$
      \EndIf
      \State $\mathit{change\_max = \max\limits_{\forall\,\textrm{neighbors}\,\tilde{a}_n}\big\{change_n\big\} }$
      \If { \newline
      $\big(\mathit{my\_term'\_counter} = \mathit{max\_distance}\big) \; \vee$
      \newline
      \mbox{$\big(change\_max - \mathit{my\_term'\_counter} > Thresh_{early\_term}\big)$} \label{alg:mgm_p2:early_termination} }
        \State \textsc{terminate}
      \EndIf
      \ForEach {neighboring agent $\tilde{a}_n$}
        \State \textbf{send} (\textbf{ok?},\enspace $\tilde{a}_l$,\enspace $\mathit{v_l}$,\enspace $r_l$, \enspace $change_l$) to $\tilde{a}_n$
      \EndFor
    \EndProcedure
  \end{algorithmic}
\end{algorithm}

\subsection{RESMGM: Heuristics and Optimization of MGM}
\label{subsec:heuristics}
In the following, we describe a set of heuristics that we apply in order to improve the scalability of the MGM algorithm.
These improvements are specific to solving constraints for manycore resource management.
We therefore refer to this improved version of MGM as \emph{Resource Management MGM} (RESMGM).
The heuristics presented below improve execution time, solution quality, or both.

\textbf{Smart initialization heuristic}:
The resource management system has to solve a new DCOP every time any application
changes its resource demands.
We initialize the a priori agents with their actual resource assignment from the
last DCOP optimization and the new agent's variables with $\textrm{FREE}$ after
the first optimization round.
In the very first DCOP optimization, when no prior resource allocation exists,
every agent takes a value that minimizes its local cost function without
taking any other agent into account.
By contrast, the original MGM implementation initializes all variables with
random values, as starting with `less conflicting' assignments reduces the effort of the initial
conflict resolution.

\textbf{Early termination heuristic}:
In contrast to the original version of MGM, we employ a different technique for quick termination detection.
The original MGM algorithm iterates for a fixed amount of $\mathit{max\_distance}$ rounds, even if an optimal solution has been found already.
In general, determining, whether a partial solution will not yield any improvement in subsequent rounds, necessitates computation and communication.
We make the following observation:
If the variable assignment does not change for any agent in round
$j, 1\!\leq\!j\!\leq \mathit{max\_distance}$, it will not change for any agent in the remaining rounds.
This is due to the monotonous improvement of the solution quality of the MGM algorithm.
We extend the \emph{ok?}-message with a Boolean field that encodes whether the agent has changed its variable in the last round or whether it has observed change from any other agent's \emph{ok?}-message.
Every agent counts the rounds in which no change was observed and terminates when this counter reaches a fixed built-in threshold value that is determined for the constraint graph of the current DCOP.
It is also possible to lower the threshold value, such that MGM terminates, when `nearly all' agents do not observe any more change in their variables.

\textbf{Early termination in local search}:
We formulate the DCOP as a minimization problem with a minimum constraint cost of zero.
If a variable assignment with zero constraint cost is found, we can abort the search without having to iterate over the entire domain of a variable.
Often variables have an optimal assignment already before the start of the optimization.
This idea can be extended.
Initially, the agents calculate the cost for giving up a resource (\emph{loss}) and the \emph{ok?}-message is extended with a field for this value.
Upon receiving an \emph{ok?}-message, agents store this value for every variable in a hash map.
The agents will then assign their variables to resources of other agents first, such that the other agent has a minimal loss, while maximizing its own decrease in solution cost (\emph{gain}).

\textbf{Incorporating system information}:
This heuristic exploits information about the hardware platform in order to solve $C_\mathrm{TileSharing}$ more efficiently.
If $C_\mathrm{TileSharing}$ applies, the search strategy is changed from iterating over individual cores to iterating over individual tiles.
Initially, the agent has to find out for every variable, if $C_\mathrm{TileSharing}$ applies to it.
The implementation of the heuristic is similar to the previous one, with the notable difference that loss and gain is accumulated for entire tiles.
The heuristic allows to avoid checking resources that are on non-free tiles.

\textbf{Thinking globally}:
In the original implementation of MGM, agents optimize their cost function locally without taking the cost of other agents into account.
This heuristic allows agents to consider loss and gain of both agents in a mutual swap of resources.
Due to the extended \emph{ok?}-message, the agent can compare its gain with the potential loss of another agent.
The agent achieves this by temporarily changing its own local view to pretend a resource is not allocated.
This heuristic does not improve execution time, but it improves the solution quality, especially in heavily loaded systems.

\textbf{Multiple variable changes}:
The original implementation finds among all neighbors the variable with the
maximum local gain and only this variable is allowed to change.
Instead, this heuristic allows to change the assignment of an arbitrary amount of
variables as long as this does not cause any conflict.
An arbitrary number of variables of one agent can be changed as long as this reduces the local cost monotonously.
The heuristic decreases the amount of rounds required by the algorithm.

\textbf{Reduced field-of-view heuristic}:
If the a priori agents aim to improve their resource assignment during the
optimization too, the amount of constraints can be reduced significantly by
letting the a priori agents choose from PEs only, that are geometrically
close to PEs on the chip, that the priori agents have claimed already.
This means that the a priori agents have a limited \emph{field of view}.
By limiting the field of view to a fixed number of tiles, the maximum amount of
constraints, and thereby conflicts, is limited to a constant and is no
longer affected by the size of the manycore system and the total amount of
agents.

\begin{figure}
	\centering
	\includegraphics[width=\columnwidth]{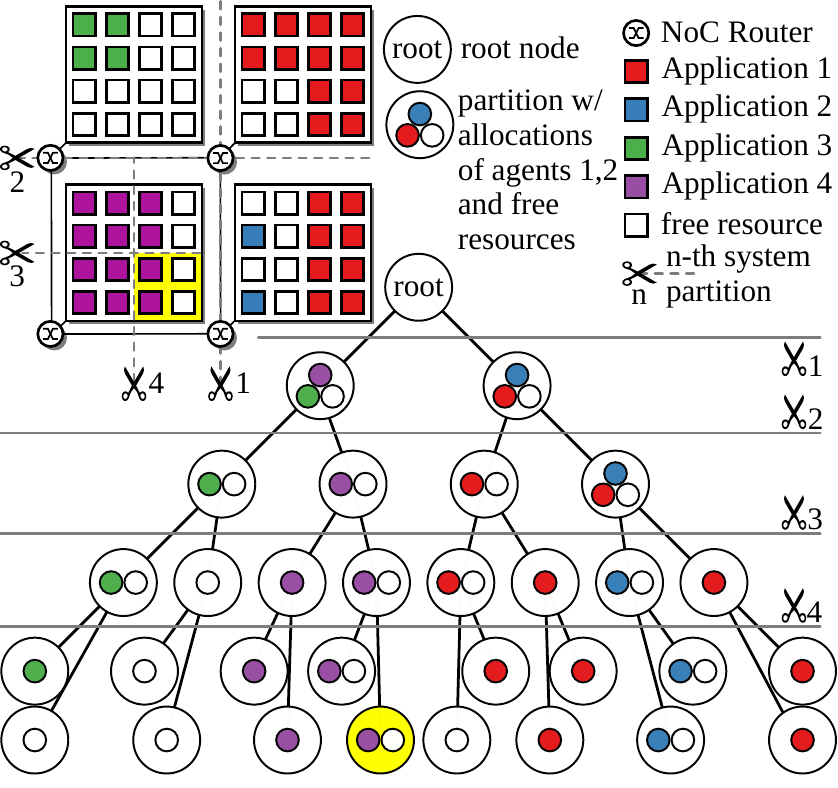}
	\caption{
		Example for resource allocation of several agents (top left).
		k-d tree over resource allocation (bottom).
		A partition and the corresponding leaf in the k-d tree are highlighted.}
	\label{fig:kdtree}
\end{figure}

\textbf{Scalability optimizations:}
Initial tests show that the large number of neighboring agents affect the scalability of RESMGM noticeably.
We therefore limit the amount of agents that can participate in the negotiation and propose a heuristic that allows to select a reasonable subset of agents by augmenting RESMGM with a \emph{k-d tree} data structure.
The manycore system is logically partitioned into partitions that group similar types and amounts of computing resources.
In the simplest case of a homogeneous system (see \cref{fig:kdtree}) this results in a binary space partition.
The levels of the tree correspond to an either vertical or horizontal cut of the manycore system as illustrated in \cref{fig:kdtree}.
There is a fixed number of levels in the tree, which is predetermined for the given manycore system.
The nodes (except for root) maintain a list of agents that have at least one resource assigned to them in the given cut.
This tree has to be updated during the execution of the resource management.
The nodes are stored in the scratchpad memory of individual cores that are associated with the cut.
When a new application requests a resource allocation, its constraint is scanned, \eg, for the requested resource types.
For example, if a constraint does not request any other core types than regular PEs, a list of agents for the negotiation is assembled from the k-d tree omitting cuts that have only non-regular resources (\eg reconfigurable PEs, see \cref{eqn:het_types}).
If a constraint involves a TileSharing constraint, the tree is scanned for a partition that contains only resources of a given agent and free resources.

\section{Evaluations}
\label{sec:evaluations}

\subsection{Experimental Setup}
We evaluated the presented method using the \emph{Sniper Multi-core Simulator} \cite{sniper}, which is a fast, accurate and parallel simulation infrastructure for x86 architectures.
We implemented our proposed RESMGM and the approach presented by \cite{anagnostopoulos_distrmclone} as POSIX threads applications that can be run in Sniper.

\begin{figure*}[h!t]
  \centering
  \includegraphics [width=\textwidth,trim={0cm 0cm 0cm 0.0cm},clip]
    {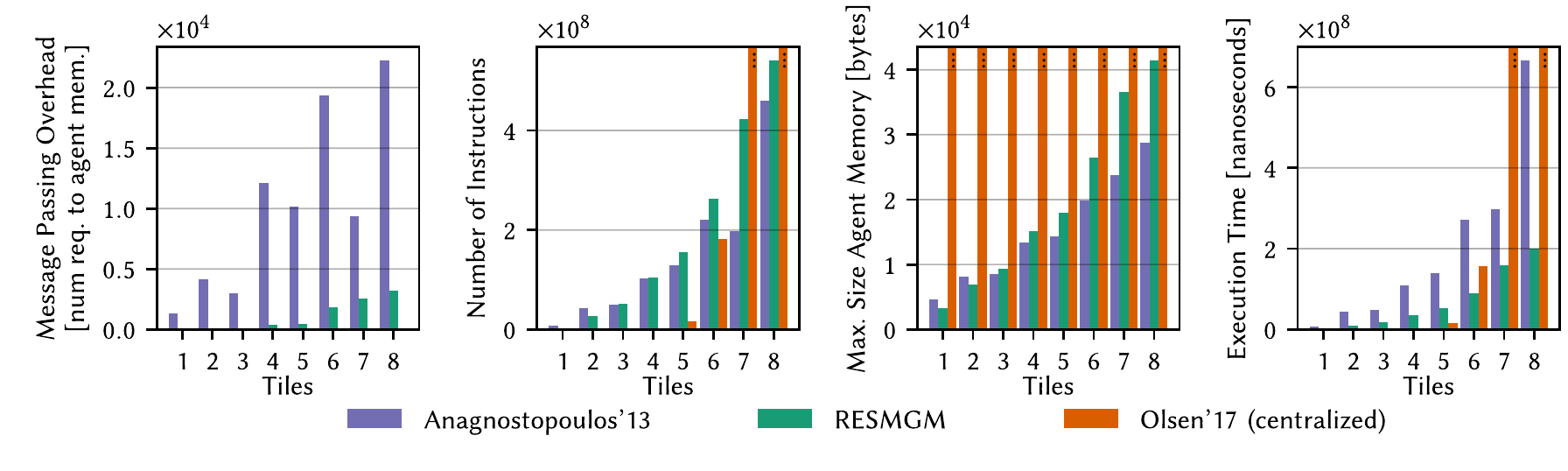}
  \vspace{-3.0em}
  \caption{
    Overheads of different resource management schemes for increasing number of tiles at constant system utilization.
    The overheads in the number of instructions and maximum size of used agent memory of Anagnostopoulos'13~\cite{anagnostopoulos_distrmclone} and RESMGM are comparable.
    The overheads in message passing overhead and execution time of RESMGM are significantly smaller than \cite{anagnostopoulos_distrmclone}.
  }
  \label{fig:var_dom_kdtree.pdf}
\vspace{-1.0em}
\end{figure*}

Sniper is configured with a configuration file that is derived from the many-core configuration which is shipped with releases of Sniper.
This configuration is loosely inspired by the Intel Xeon Phi and models a many-core system that features many small cores with private L1 and L2 caches and directory-based cache coherency.
We use this configuration for our simulations and modify it in order to model a tile-based architecture (see \cref{fig:tilearch}) in Sniper.
For a tile-based architecture we need to model and measure messages that are sent between tiles, \ie, communication between agents.
Therefore, we extend the model and introduce large private L3 caches per core.
If used properly, read and write access to the L3 caches can be regarded as a measure of the message passing overhead.

At initialization of an agent object on a specific core, an isolated area of memory (``agent memory'') is allocated for the agent such that it resides entirely in the simulated L3 cache.
Agent memory is only directly accessible by a specific agent and is entirely self-contained, \ie, not only the actual data is stored in it, but also the necessary bookkeeping.
The agent memory stores the agent's message queue, its local view, and every other piece of private data.
Before starting the resource management algorithm, agents load their agent memory into their private cache simulated in Sniper.
Each agent runs on its own simulated core in Sniper by creating a separate POSIX thread and pinning it to a unique core.

Communication between agents occurs via simple message passing primitives.
Messages are constructed in the agent memory and appended to another agent's message queue.
In order to read a message, the receiving agent retrieves a pointer from the message in the message queue and reads the content that is located in the sending agent's agent memory.
Subsequent cache-coherency traffic between cores is therefore directly related to read accesses of remote agents to another agent's private cache.
This can be exploited to measure communication overhead of the algorithms.
Note that the simulations only execute the resource management algorithm to ensure that the message passing evaluation is not disturbed by application activity.

In addition to our proposed RESMGM, we implemented the resource management scheme introduced by \cite{anagnostopoulos_distrmclone} to compare against RESMGM.
Ref.~\cite{anagnostopoulos_distrmclone} introduces a workload-aware distributed runtime framework for malleable applications, \ie, applications that support changing their resources at runtime.
In our experiments, we generate random constraints that are inspired by, but not limited to the constraints required for the multigrid solver application introduced in \cite{buchwald2015malleable}.
The multigrid solver has a master-slave structure and a global queue of jobs.
For the individual jobs, it defines $C_\mathrm{PEQuantity}$ to request a
single-digit amount of resources, a scalability with
$C_\mathrm{Downey}$ and $C_\mathrm{TileSharing}$ for some jobs.
Each a priori agent is assigned a random set of resources, in accordance with a user-defined a priori system load and a specified maximum number of resources per individual agent.
By \emph{a priori system load $\in [0,1]$} we understand the ratio of the number of initially allocated resources over the total number of resources in the system.
The metrics we obtain from Sniper are execution time, number of executed instructions, number of requests to agent memory and maximum size of used agent memory.
The number of requests to agent memory is equal to the number of loads from remote caches necessary for cache coherency measured by the Sniper framework.
The numbers, that we report, are averaged over all cores of a system.

\begin{figure*}[h!t]
	\centering
	\includegraphics [width=\textwidth,trim={0cm 0cm 0cm 0cm},clip]
    {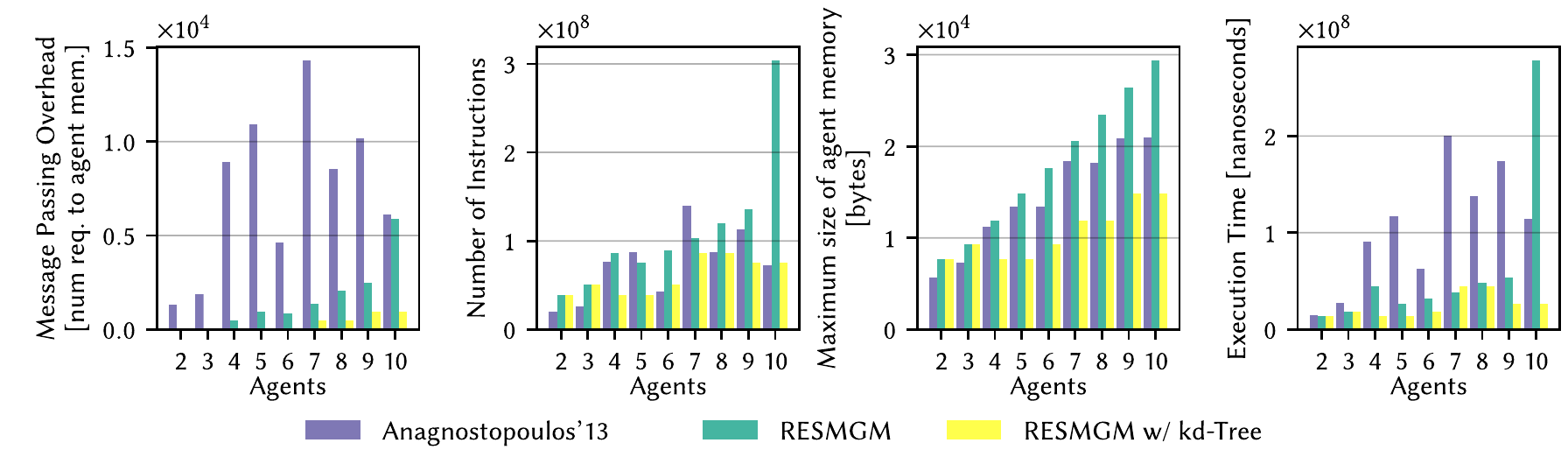}
	\vspace{-2.5em}
	\caption{Overheads of different resource management schemes for increasing number of agents at constant system size.
						The overheads in the number of instructions and maximum size of used agent memory of Anagnostopoulos'13~\cite{anagnostopoulos_distrmclone} and RESMGM are comparable.
						The overheads in message passing and execution time of RESMGM are significantly smaller than \cite{anagnostopoulos_distrmclone}.
						An exception is the case of a fully loaded system in which RESMGM needs to test a big number of possible resource assignments before finding a valid solution.
						This creates high overheads in the number of instructions and execution time.
					}
    \label{fig:var_ag_kdtree.pdf}
\end{figure*}

A limitation of the comparison between RESMGM and \cite{anagnostopoulos_distrmclone} is, that they do not support constraints, except for $C_\mathrm{Downey}$.
In order to allow for a more fair comparison, $C_\textrm{TileSharing}$ cannot be used.
Ref.\ \cite{anagnostopoulos_distrmclone} does not distinguish between different types of resources and disregards the intricate resource requirements that applications formulate with a set of constraints.
This intrinsic deficiency of \cite{anagnostopoulos_distrmclone} puts RESMGM naturally at a disadvantage in the following evaluations, since it is significantly more complex to satisfy an entire set of constraints instead of balancing different scaling properties of a number of applications.
The following experiments were conducted on a system with an Intel Core i7-6700T CPU with a clock rate of $2.80\,\mathrm{GHz}$ and $32\,\mathrm{GiB}$ DDR3 RAM running Ubuntu 17.04 (GNU/Linux 4.10.0-33-generic x86\_64).

\subsection{Simulation Results}
\subsubsection{Increasing the number of tiles}

In the first experiment, we vary the domain size, \ie, the amount of tiles in the manycore system, ranging from 1 to 8.
The a priori system load is at a constant of 40\% and the number of a priori agents is adjusted accordingly.
A priori agents are assigned up to 4 resources and the new agent requires 2 to 4 resources of any type.
The overheads measured in the experiment are shown in \cref{fig:var_dom_kdtree.pdf}.
RESMGM has a significantly smaller execution time and message passing overhead, although being at a disadvantage in the simulation setup.
The number of instructions and the maximum size of used agent memory are comparable for both approaches.
Reducing the amount of agents that participate in the negotiation with the k-d tree helps to reduce the overheads further, albeit visually it does not seem to alter the scalability of our approach itself. For small numbers of tiles, the bars in \cref{fig:var_dom_kdtree.pdf} for RESMGM w/ k-d tree are too small to be visually seen in the plot.
Note that \cite{anagnostopoulos_distrmclone} is a randomized algorithm and the results may vary depending on random seed, system state etc.

\subsubsection{Increasing the number of agents}
In the second experiment, the number of a priori agents is increased from 2 to 10 on a system with 3 tiles.
A priori agents only claim up to 2 resources, while the new agent continues to demand 2 to 4 resources of any type.
The a priori system load ranges from 11\% (with two agents) to 100\% (with ten agents).
The performed measurements are analogous to the previous experiment.
Our findings are shown in \cref{fig:var_ag_kdtree.pdf}.
Similar to the first experiment, the number of instructions and maximum size of used agent memory is comparable for RESMGM and \cite{anagnostopoulos_distrmclone}.
Message passing overhead and execution times are always smaller for RESMGM, except for the case of a highly loaded system, in which RESMGM has an execution time of more than 2$\times$ compared to \cite{anagnostopoulos_distrmclone}.
This is because the RESMGM agents need to test a big number of possible variable assignments and evaluate the corresponding constraints before finding a valid solution in a highly loaded system.
Again, for a small numbers of agents, the bars in \cref{fig:var_ag_kdtree.pdf} for RESMGM w/ k-d tree are too small to be visually seen in the plot.

\section{Conclusions}
\label{sec:conclusions}
The resources of state\-/of\-/the\-/art manycore systems are heterogeneous for efficiency reasons.
Their management, \ie, mapping application to cores, has become a key problem since workloads are increasingly complex.
We present a DCOP formulation of the manycore resource allocation problem that allows finding a resource allocation at runtime in a decentralized manner for applications that provide constraints on their resource requirements.
Our proposed local search algorithm RESMGM finds resource allocations quickly and efficiently with a high solution quality requiring a feasible computational overhead.
We demonstrate the effectiveness of our proposed approach with an evaluation using the Sniper multi\-/core simulator.
The computational overheads of the state\-/of\-/the\-/art approach \cite{anagnostopoulos_distrmclone} and RESMGM are mostly comparable, but RESMGM allows applications to express a wide range of constraints flexibly.

\section*{Acknowledgment}
Funded by the Deutsche Forschungsgemeinschaft (DFG, German Research Foundation)
 -- Project Number 146371743 -- TRR 89 Invasive Computing

\bibliographystyle{IEEEtran}
\bibliography{RESMGM_bibliography}

\begin{thebibliography}{10}
\providecommand{\url}[1]{#1}
\csname url@samestyle\endcsname
\providecommand{\newblock}{\relax}
\providecommand{\bibinfo}[2]{#2}
\providecommand{\BIBentrySTDinterwordspacing}{\spaceskip=0pt\relax}
\providecommand{\BIBentryALTinterwordstretchfactor}{4}
\providecommand{\BIBentryALTinterwordspacing}{\spaceskip=\fontdimen2\font plus
\BIBentryALTinterwordstretchfactor\fontdimen3\font minus
  \fontdimen4\font\relax}
\providecommand{\BIBforeignlanguage}[2]{{%
\expandafter\ifx\csname l@#1\endcsname\relax
\typeout{** WARNING: IEEEtran.bst: No hyphenation pattern has been}%
\typeout{** loaded for the language `#1'. Using the pattern for}%
\typeout{** the default language instead.}%
\else
\language=\csname l@#1\endcsname
\fi
#2}}
\providecommand{\BIBdecl}{\relax}
\BIBdecl

\bibitem{borkar07thousands}
S.~Borkar, ``Thousand core chips: A technology perspective,'' in \emph{Design
  Automation Conference (DAC)}, 2007, pp. 746--749.

\bibitem{IntelKnightsLanding2016}
A.~Sodani, R.~Gramunt, J.~Corbal, H.~S. Kim, K.~Vinod, S.~Chinthamani,
  S.~Hutsell, R.~Agarwal, and Y.~C. Liu, ``{Knights Landing}: Second-generation
  {Intel Xeon Phi} product,'' \emph{IEEE Micro}, vol.~36, no.~2, pp. 34--46,
  2016.

\bibitem{ebi_tape}
T.~Ebi, M.~A. Al~Faruque, and J.~Henkel, ``{TAPE}: Thermal-aware agent-based
  power economy for multi/many-core architectures,'' in \emph{Intl. Conf. on
  Computer-Aided Design (ICCAD)}, 2009, pp. 302--309.

\bibitem{cao2002arms}
J.~Cao, S.~A. Jarvis, S.~Saini, D.~J. Kerbyson, and G.~R. Nudd, ``{ARMS}: An
  agent-based resource management sys. for grid comput.'' \emph{Scientific
  Programming}, vol.~10, no.~2, pp. 135--148, 2002.

\bibitem{Ge:2010:DTM:1837274.1837417}
Y.~Ge, P.~Malani, and Q.~Qiu, ``Distributed task migration for therm. manag. in
  many-core systems,'' in \emph{Design Automation Conference (DAC)}, 2010, pp.
  579--584.

\bibitem{kobbe_distrm}
S.~Kobbe, L.~Bauer, D.~Lohmann, W.~Schr\"{o}der-Preikschat, and J.~Henkel,
  ``{DistRM}: Distributed resource management for on-chip many-core systems,''
  in \emph{Intl. Conf. on Hardware/Software Codesign and System Synthesis
  (CODES+ISSS)}, 2011, pp. 119--128.

\bibitem{hannig_resource_aware}
F.~Hannig, S.~Roloff, G.~Snelting, J.~Teich, and A.~Zwinkau, ``Resource-aware
  programming and simulation of {MPSoC} architectures through extension of
  {X10},'' in \emph{Intl. Workshop on Software and Compilers for Emb. Sys.
  (SCOPES)}, 2011, pp. 48--55.

\bibitem{bungartz_multigrid}
H.-J. Bungartz, C.~Riesinger, M.~Schreiber, G.~Snelting, and A.~Zwinkau,
  ``Invasive computing in {HPC} with {X10},'' in \emph{SIGPLAN X10 Workshop
  (X10)}, 2013, pp. 12--19.

\bibitem{singh_review}
A.~K. Singh, M.~Shafique, A.~Kumar, and J.~Henkel, ``Mapping on multi/many-core
  sys.: Surv. of curr. and emerg. trends,'' in \emph{Design Automat. Conf.
  (DAC)}, 2013, pp. 1:1--1:10.

\bibitem{Singh:2017:SCS:3071073.3057267}
A.~K. Singh, P.~Dziurzanski, H.~R. Mendis, and L.~S. Indrusiak, ``A survey and
  comparative study of hard and soft real-time dynamic resource allocation
  strategies for multi-/many-core systems,'' \emph{ACM Computing Survey},
  vol.~50, no.~2, pp. 24:1--24:40, 2017.

\bibitem{Fattah:2013:SHC:2463209.2488782}
M.~Fattah, M.~Daneshtalab, P.~Liljeberg, and J.~Plosila, ``Smart hill climb.
  for agile dyn. mapping in many-core systems,'' in \emph{Design Automation
  Conference (DAC)}, 2013, pp. 39:1--39:6.

\bibitem{6654651}
G.~Castilhos, M.~Mandelli, G.~Madalozzo, and F.~Moraes, ``Distributed res.
  manag. in {NoC}-based {MPSoCs} with dyn. cluster sizes,'' in \emph{Symposium
  on VLSI (ISVLSI)}, 2013, pp. 153--158.

\bibitem{anagnostopoulos_distrmclone}
I.~Anagnostopoulos, V.~Tsoutsouras, A.~Bartzas, and D.~Soudris, ``Distributed
  run-time resource management for malleable applications on many-core
  platforms,'' in \emph{Design Automation Conference (DAC)}, 2013, pp.
  168:1--168:6.

\bibitem{melissaris_agora}
T.~Melissaris, I.~Anagnostopoulos, D.~Soudris, and D.~Reisis, ``Agora: Agent
  and market-based resource management for many-core systems,'' in \emph{Intl.
  Conference on Electronics, Circuits and Systems (ICECS)}, 2016.

\bibitem{wildermann_multi_objective}
S.~Wildermann, M.~Gla\ss, and J.~Teich, ``Multi-objective distributed run-time
  resource management for many-cores,'' in \emph{Conference on Design,
  Automation \& Test in Europe (DATE)}, 2014, pp. 221:1--221:6.

\bibitem{olsen_res_mgmt}
D.~Olsen and I.~Anagnostopoulos, ``Performance-aware resource management of
  multi-threaded applications on many-core systems,'' in \emph{Great Lakes
  Symposium on VLSI}, 2017.

\bibitem{mgm}
R.~T. Maheswaran, J.~P. Pearce, and M.~Tambe, ``Distributed algorithms for
  {DCOP}: A graphical-game-based approach.'' in \emph{Intl. Conf. on Parallel
  and Distr. Computing and Comm. Systems (ISCA PDCS)}, 2004, pp. 432--439.

\bibitem{yeoh08bnbadopt}
W.~Yeoh, A.~Felner, and S.~Koenig, ``{BnB-ADOPT}: An asynchronous
  branch-and-bound {DCOP} algorithm,'' in \emph{Intl. Joint Conf. on Autonomous
  Agents and Multiagent Systems - Volume 2 (AAMAS)}, 2008, pp. 591--598.

\bibitem{downey}
A.~B. Downey, ``A parallel workload model and its implications for processor
  allocation,'' \emph{Cluster Comput.}, vol.~1, no.~1, pp. 133--145, 1998.

\bibitem{dba}
M.~Yokoo and K.~Hirayama, ``Distributed breakout algorithm for solving
  distributed constraint,'' in \emph{Intl. Conf. on Multiagent Systems}, 1996,
  pp. 401--408.

\bibitem{sniper}
T.~E. Carlson, W.~Heirmant, and L.~Eeckhout, ``Sniper: Exploring the level of
  abstraction for scalable and accurate parallel multi-core simulation,'' in
  \emph{Intl. Conf. for High Perf. Comput., Network., Storage and Analysis
  (SC)}, 2011.

\bibitem{buchwald2015malleable}
S.~Buchwald, M.~Mohr, and A.~Zwinkau, ``Malleable invasive applications,'' in
  \emph{Software Engineering (Workshops)}, 2015, pp. 123--126.

\end{thebibliography}

\newpage

\begin{IEEEbiography}
  [{ \includegraphics[width=1in,height=1.25in,clip,keepaspectratio]{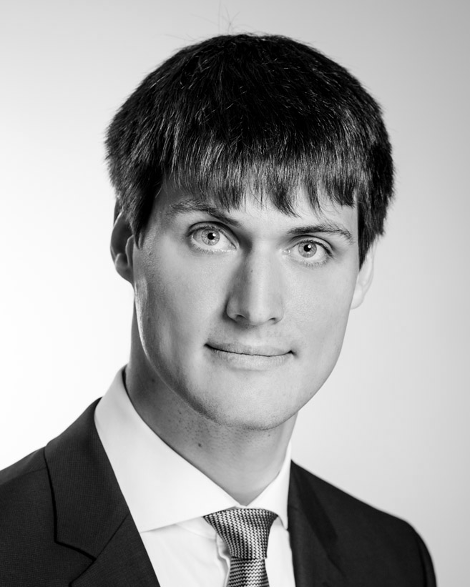} }]
  {Volker Wenzel}
  worked under the supervision of Prof.\ Jörg Henkel at the Chair for Embedded Systems (CES), Karlsruhe Institute of Technology, Germany from May 2013 until Oct. 2018.
\end{IEEEbiography}

\begin{IEEEbiography}
  [{ \includegraphics[width=1in,height=1.25in,clip,keepaspectratio]{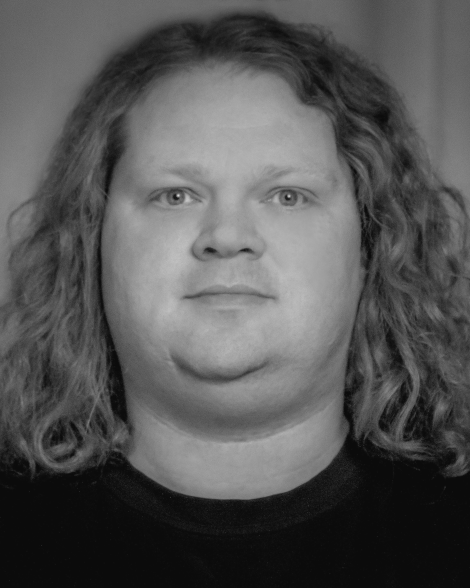} }]
  {Lars Bauer}
  received the M.Sc.\ and Ph.D.\ degrees in computer science from the University of Karlsruhe, Germany, in 2004 and 2009.
  He is currently a research group leader and lecturer at the Chair for Embedded Systems (CES) at the Karlsruhe Institute of Technology (KIT).
  Dr.\ Bauer received two dissertation awards (EDAA and FZI), two best paper awards (AHS'11 and DATE'08) and several nominations.
  His research interests include architectures and management for adaptive multi-/manycore systems.
\end{IEEEbiography}

\begin{IEEEbiography}
  [{ \includegraphics[width=1in,height=1.25in,clip,keepaspectratio]{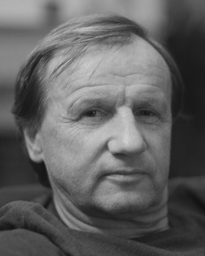} }]
  {Wolfgang Schr\"oder-Preikschat}
  studied computer science at the Technical University of Berlin, Germany, where he also took his doctor's degree and venia legendi.
  After a long-term period of extramural research at GMD/FhG, Berlin, Germany, and ICSI, Berkeley, USA, he became full professor for computer science at University of Potsdam, Magdeburg, and Erlangen, Germany, likewise.
  Dr. Schr\"oder-Preikschat is member of ACM, EuroSys, GI, IEEE, and USENIX.
  His main research interest is on resource-aware (parallel) operating systems, notably process coordination, especially as to time/energy-dependable application and problem domains.
\end{IEEEbiography}

\begin{IEEEbiography}
  [{ \includegraphics[width=1in,height=1.25in,clip,keepaspectratio]{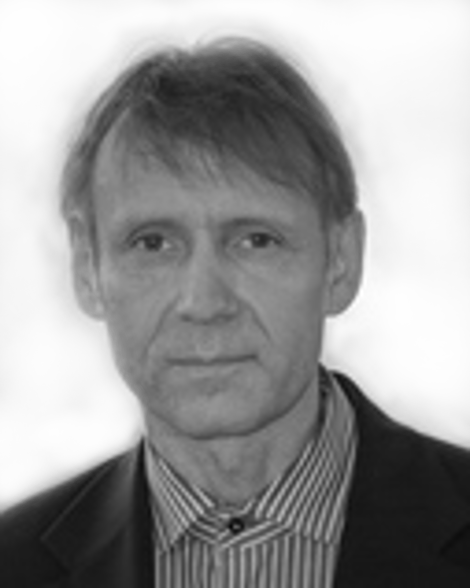} }]
  {J\"org Henkel}
  received the Diploma and Ph.D.\ (summa cum laude) degree from the Technical University of Braunschweig.
  He is currently the Chair Professor of embedded systems with the Karlsruhe Institute of Technology.
  Before that he was a Research Staff Member with NEC Laboratories, Princeton, NJ, USA.
  His research interest includes co-design for embedded hardware/software systems with respect to power security and means of embedded machine learning.
  He has led several conferences as a General Chair including ICCAD and ESWeek, and is currently DAC Vice Chair.
  He serves as a steering committee chair/member for leading conferences and journals for embedded and cyber-physical systems.
  He has coordinated the DFG Program SPP 1500 ``Dependable Embedded Systems'' and is a Site Coordinator of the DFG TR89 Collaborative Research Center on ``Invasive Computing''.
  He is the Chairman of the IEEE Computer Society, Germany Chapter.
  He has received six best paper awards throughout his career from, among others, ICCAD, ESWeek, and DATE.
  For two consecutive terms each, he served as the Editor-in-Chief for both the ACM Transactions on Embedded Computing Systems and the IEEE Design \& Test magazine.
  He is the Vice President for Publications at IEEE CEDA and a Fellow of the IEEE.
\end{IEEEbiography}

\vfill

\end{document}